\documentclass[12pt]{revtex4}
\usepackage{graphicx}
\usepackage{amsmath}

\begin{document}

\title{Conformational changes in glycine tri- and hexapeptide}

\author{Alexander V. Yakubovitch,
        Ilia A. Solov'yov\footnote{On leave from the A.F. Ioffe Institute, St. Petersburg, Russia.
E-mail: ilia@th.physik.uni-frankfurt.de},
        Andrey V. Solov'yov\footnote{On leave from the A.F. Ioffe Institute, St. Petersburg, Russia.
E-mail: solovyov@fias.uni-frankfurt.de},
        and Walter Greiner}

\address{Frankfurt Institute for Advanced Studies,
Max von Laue Str. 1, 60438 Frankfurt am Main, Germany}

\begin{abstract}
We have investigated the potential energy surfaces for glycine
chains consisting of three and six amino acids. For these molecules
we have calculated potential energy surfaces as a function of the
Ramachandran angles $\varphi$ and $\psi$, which are widely used for
the characterization of the polypeptide chains. These particular
degrees of freedom are essential for the characterization of
proteins folding process. Calculations have been carried out within
{\it ab initio} theoretical framework based on the density
functional theory and accounting for all the electrons in the
system. We have determined stable conformations and calculated the
energy barriers for transitions between them. Using a thermodynamic
approach, we have estimated the times of the characteristic
transitions between these conformations. The results of our
calculations have been compared with those obtained by other
theoretical methods and with the available experimental data
extracted from the Protein Data Base. This comparison demonstrates a
reasonable correspondence of the most prominent minima on the
calculated potential energy surfaces to the experimentally measured
angles $\varphi$ and $\psi$ for the glycine chains appearing in
native proteins. We have also investigated the influence of the
secondary structure of polypeptide chains on the formation of the
potential energy landscape. This analysis has been performed for the
sheet and the helix conformations of chains of six amino acids.
\end{abstract}

\maketitle

\section{Introduction}

It is well known that proteins consist of amino acids whose number
may vary in the range from hundreds up to tens of thousands. Small
fragments of proteins are usually called polypeptide chains or
polypeptides. This work is devoted to a study of the conformational
properties of glycine polypeptide chains.

Since recently, it became possible to study experimentally small
fragments of proteins and polypeptides in the gas phase with the use
of the MALDI mass spectroscopy \cite{Karas88,Karas00,Karas03,Wind04}
and the ESI mass spectroscopy \cite{Fenn89,Hvelplund04}. From
theoretical view point, investigation of small polypeptides is of
significant interest because they can be treated by means of {\it ab
initio} methods allowing accurate comparison of theoretical
predictions with experiment. The results of {\it ab initio}
calculations can be then utilized for the development of model
approaches applicable for the description of larger and more complex
protein structures.

Polypeptides are characterized by the primary and the secondary
structure \cite{Ptizin_book,Muelberg_book,Protbase,Rubin04}.
Different geometrical configurations of a polypeptide are often
called as the conformations. One can expect that chemical and
physical properties of various conformations of complex molecules
might differ significantly. The number of various conformations
(isomeric states) grows rapidly with the growth of a system size.
Thus, a search for the most stable conformations becomes an
increasingly difficult problem for large molecules. With the help of
the NMR spectroscopy and the X-rays diffraction analysis it has been
shown \cite{Protbase}, that the sheet and the helix structures are
the most prominent elements of the protein secondary structure.

The main difference between the sheet and the helix structures is
due to the difference of the dihedral angles formed by the atoms of
the polypeptide chains in the two cases. These degrees of freedom
are responsible for the transition of the molecule from one
conformation to another. By increasing the temperature of the
system, the degrees of freedom responsible for twisting of the
polypeptide chain can be activated. The study of this transition and
evaluation of its characteristic duration are of significant
interest, because this problem is closely related to one of the most
intriguing problems of the protein physics - the protein folding. To
study this transition it is necessary to investigate the potential
energy surface of amino acid chains with respect to their twisting.
Besides the protein folding, the potential energy landscapes of
polypeptides carry a lot of detail and useful information about the
structure of these molecules.

In the present paper we have studied the potential energy surface
for small glycine chains. These molecules were chosen because they
are often present in native proteins as fragments, and also because
they allow {\it ab initio} theoretical treatment due to their
relatively small size.

Previously, only glycine and alanine dipeptides were studied in
detail. Sometimes their analogues were used to reduce the
computational costs (for example
(S)-$\alpha$-(formylamino)propanamide). In refs.
\cite{Head-Gordon91,Gould94,Zhi-Xiang04} alanine and glycine
dipeptides were investigated within the Hartree-Fock theory. In
these papers the potential energy surfaces were calculated versus
the twisting angles of the molecules. Different stable conformations
of the dipeptides, corresponding to different molecular
conformations, were determined. Each stable conformation of the
molecule was additionally studied on the basis of the perturbation
theory, which takes into account many-electron correlations in the
system. In refs.
\cite{Percel03,Hudaky04,Improta04,Vargas02,Kashner98,Salahub01}
different conformations and their energies were determined within
the framework of the density functional theory. In ref.
\cite{Salahub01} dynamics of the alanine dipeptide analog was
discussed and time of the transition between the two conformations
of the alanine dipeptide was found. In \cite{Hobza00} the glycine
molecule was studied with accounting for the many electron
correlations within the many-body perturbation theory. In this work
the most important glycine isomers and the frequencies of the normal
vibration modes were determined.

A number of papers were devoted to the study of tripeptides. In
refs. \cite{Woutersen01_1,Woutersen02,Stock02,Stock03,Stock03_1}
dynamics of the alanine and glycine tripeptides was studied by means
of classical molecular dynamics and  with the use of semi-empirical
potentials (such as GROMOS, CHARMM and AMBER). In ref.
\cite{Torii98} within the framework of the Hartree-Fock theory
several stable conformations of alanine and glycine tripeptides were
found. In ref. \cite{Stenner01} the Raman and IR spectra for alanine
and glycine tripeptides were measured in neutral, acidy and alkali
environments.

Polypeptides have been studied less. We are aware of only several
related papers. In particular, stable conformations of neutral and
charged alanine hexapeptides were obtained with the use of empirical
potentials and discussed in ref. \cite{Levy}. Experimental NMR study
of various conformations of alanine heptapeptides at different
temperatures was carried out in ref. \cite{Shi02}. In ref.
\cite{Garcia03} with the use of empirical molecular dynamics based
on Monte-Carlo methods, a polypeptide consisting of 21 amino acids
was described.

In the present paper we have performed an {\it ab initio}
calculation of the multidimensional potential energy surface for the
glycine polypeptide chains consisting of three and six amino acids.
The potential energy surface versus twisting degrees of freedom of
the polypeptide chain has been calculated. The calculations have
been performed within {\it ab initio} theoretical framework based on
the density functional theory (DFT) accounting for all the electrons
in the system. Previously, this kind of calculations were performed
only for dipeptides (see, e.g.,
\cite{Head-Gordon91,Gould94,Salahub01}). For larger molecules, only
a few conformations were considered (see citations above). We have
calculated the energy barriers for the transitions between different
molecular conformations and determined the energetically most
favorable ones. Using a thermodynamic approach, we have estimated
times of the characteristic transitions between the most
energetically favorable conformations. The results of our
calculation have been compared with other theoretical simulations
and with the available experimental data. We have also analysed how
the secondary structure of polypeptide chains influences the
potential energy landscapes. In particular, the role of the
secondary structure in the formation of stable conformations of the
chains of six amino acids being in the sheet and in the helix
conformations has been elucidated. Some preliminary results of our
work were published as electronic preprints
\cite{twisting_preprint,fission_preprint}.

Our paper is organized as follows.  In section \ref{theory} we give
a brief overview of theoretical methods used in our work. In section
\ref{results} we present and discuss the results of our
computations. In section \ref{conclusions} we draw a conclusion to
this paper. The atomic system of units, $|e|=m_e=\hbar=1$, is used
throughout the paper unless other units are indicated.

\section{Theoretical methods}
\label{theory}

In the present paper we study the multidimensional potential energy
surfaces for glycine polypeptides within the framework of the
density functional theory. The potential energy surfaces are
multidimensional functions of atomic coordinates. In our work the
potential energy surfaces were considered as a function of the
dihedral angles formed by the atoms of the polypeptide chain. For
this calculation the Born-Oppenheimer approximation allowing to
separate the motion of the electronic and ionic subsystems is used.

The density functional theory (DFT) is a common tool for the
calculation of properties of quantum many body systems in which many
electron correlations play an important role. The DFT formalism is
well known and can be found in many textbooks (see e.g.
\cite{LesHouches,ISACC2003}). Therefore in our work we present only
the basic equations and ideas of this method.

Electronic wave functions and energy levels within the framework of
DFT are obtained from the Kohn-Sham equations, which read as (see
e.g. \cite{LesHouches,ISACC2003}):

\begin{equation}
\left( \frac{\hat p^2}2+U_{ions}+V_{H}+V_{xc}\right) \psi_i
=\varepsilon _i \psi _i,\ \ \ \ \ i=1...N
\end{equation}

\noindent where the first term represents the kinetic energy of the
$i$-th electron with the wavefunction $\psi_i$ and the energy
$\varepsilon_i$, $U_{ions}$ describes the electron attraction to the
ionic centers, $V_{H}$ is the Hartree part of the interelectronic
interaction \cite{Lindgren}, $V_{xc}$ is the local
exchange-correlation potential.

The exchange-correlation potential is defined as a functional
derivative of the exchange-correlation energy functional:

\begin{equation}
V_{xc}=\frac{\delta E_{xc}[\rho]}{\delta \rho(\vec r)},
\label{Vxc}
\end{equation}

\noindent Equation (\ref{Vxc}) is exact and follows from the
Hohenberg theory \cite{Hohenberg64}. However, no unique potential
$E_{xc}$, universally applicable for different systems and
conditions, has been found so far.

Approximate functionals employed by the DFT usually partition the
exchange-correlation energy into two parts, referred to as the {\it
exchange} and the {\it correlation} terms:

\begin{equation}
E_{xc}[\rho]= E_x(\rho)+E_c(\rho)
\label{ex_core}
\end{equation}

\noindent Both terms are the functionals of the electron density,
which can be of two distinctly different types: either a {\it local}
functional depending only on the electron density $\rho$ or a {\it
gradient-corrected} functionals depending on both $\rho$ and its
gradient, ${\bf \nabla} \rho$. A variety of exchange correlation
functionals can be found in literature. In our work we have used the
hybrid Becke-type three-parameter exchange functional \cite{Becke88}
paired with the gradient-corrected Lee, Yang and Parr correlation
functional ($B3LYP$) \cite{LYP,Parr-book}.

\section{Results and Discussion}
\label{results}

\subsection{Determination of the polypeptide twisting degrees of freedom}

\begin{figure}[h]
\includegraphics[scale=0.6,clip]{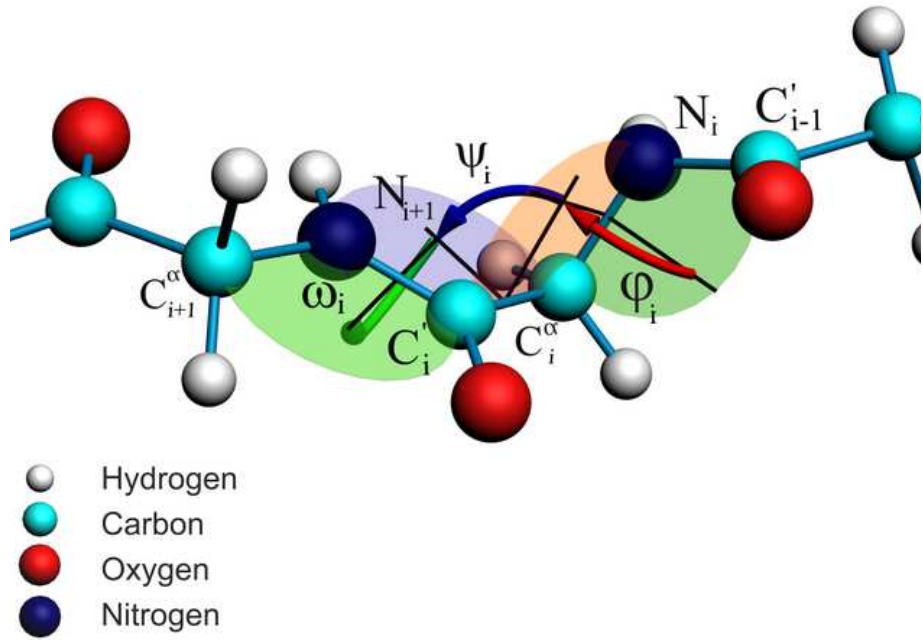}
\caption{Dihedral angles $\varphi$ and $\psi$ used to characterize
the potential energy surface of the polypeptide chain.}
\label{angles_def}
\end{figure}

In this section we present the potential energy surfaces for the
glycine polypeptide chains calculated versus dihedral angles
$\varphi$ and $\psi$ defined in figure \ref{angles_def}. In
particular, we focus on the chains consisting of three and six amino
acids.

Both angles are defined by the four neighboring atoms in the
polypeptide chain. The angle $\varphi_i$ is defined as the dihedral
angle between the planes formed by the atoms
($C_{i-1}^{'}-N_{i}-C_i^{\alpha}$) and
($N_{i}-C_i^{\alpha}-C_i^{'}$). The angle $\psi_i$ is defined as the
dihedral angle between ($N_{i}-C_{i}^{\alpha}-C_i^{'}$) and
($C_{i}^{\alpha}-C_i^{'}-N_{i+1}$) planes. Beside the angles
$\varphi_i$ and $\psi_i$ there is an angle $\omega_i$, which is
defined as the dihedral angle between the
($C_{i}^{\alpha}-C_i^{'}-N_{i+1}$) and
($C_{i}^{'}-N_{i+1}-C_{i+1}^{\alpha}$) planes. The atoms are
numbered from the $NH_2-$ terminal of the polypeptide. The angles
$\varphi_i$, $\psi_i$ and $\omega_i$ take all possible values within
the interval [$-180^{\circ}$;$180^{\circ}$]. For the unambiguous
definition we count the angles $\varphi_i$, $\psi_i$ and $\omega_i$
clockwise, if one looks on the molecule from its $NH_2-$ terminal
(see fig. \ref{angles_def}). This way of angle counting is the most
commonly used \cite{Rubin04}.

The angles $\varphi_i$ and $\psi_i$ can be defined for any amino
acid in the chain except for the first and the last ones. Below we
omit the subscripts and consider angles $\varphi$ and $\psi$ for the
middle amino acid of the polypeptide.

\subsection{Optimized geometries of glycine polypeptides}

In order to study the twisting of a polypeptide chain one needs
first to define its initial structure. Although the number of its
conformations increases with the growth of the molecule size, there
are certain types of polypeptide structure, namely the sheet and the
helix conformations, which are the most typical. In the present
paper we have investigated twisting of the polypeptide chains of the
sheet and the helix conformation. By varying the angles $\varphi$
and $\psi$ in the central amino acid one can create the structure of
the polypeptide differing significantly from the pure sheet or helix
conformations. If the structure of a polypeptide can be transformed
to a helix or a sheet one by a trivial variation of $\varphi$ and
$\psi$, such polypeptides are referred below as belonging to the
group of the helix or the sheet structure respectively.

\begin{figure}[h]
\includegraphics[scale=0.8,clip]{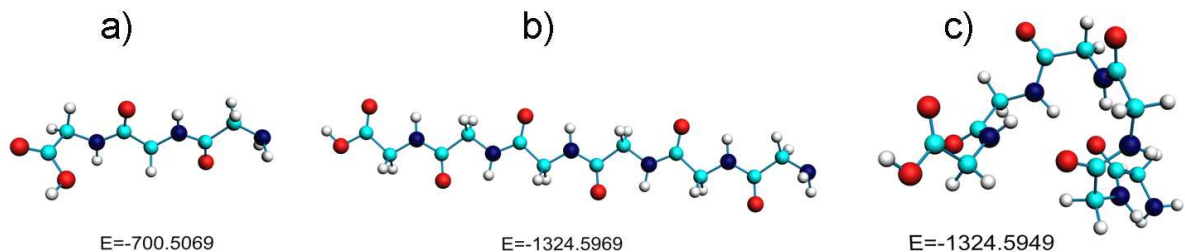}
\caption{Optimized geometries of glycine polypeptide chains
calculated by the B3LYP/6-31++G(d,p) method: a) Glycine tripeptide;
b) Glycine hexapeptide (sheet conformation); c) Glycine hexapeptide
(helix conformation).} \label{stable_geom}
\end{figure}

In figure \ref{stable_geom} we present the optimized geometries of
glycine polypeptide chains that have been used for the exploration
of the potential energy surfaces. All geometries have been optimized
with the use of the B3LYP functional. Figure \ref{stable_geom}a
shows the glycine tripeptide structure. In the present work we
choose in the sheet conformation, because the tripeptide is too
short to form the helix conformation. Figures \ref{stable_geom}b and
\ref{stable_geom}c show glycine hexapeptide in the sheet and the
helix conformations respectively. The total energies (in atomic
units) of the molecules are given below the images.

\subsection{Potential energy surface for glycine tripeptide}

\begin{figure}[h]
\includegraphics[scale=0.8,clip]{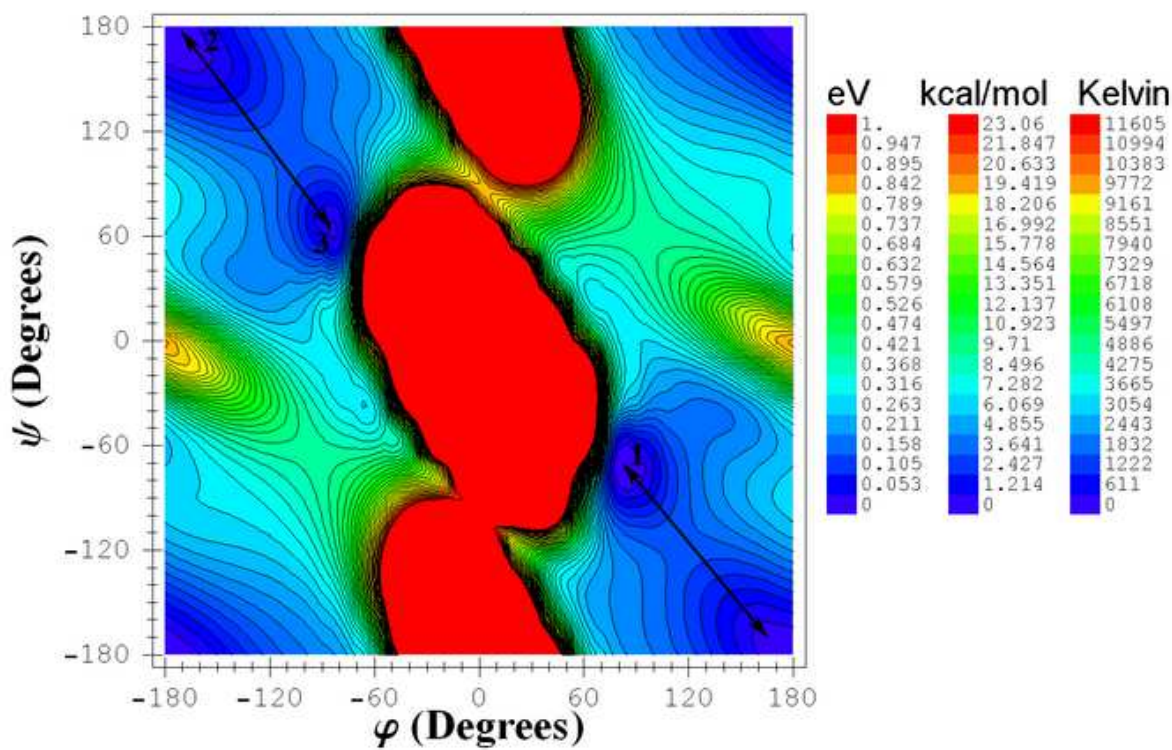}
\caption{Potential energy surface for the glycine tripeptide
calculated by the B3LYP/6-31G(2d,p) method. Energies are given in
eV, kcal/mol and Kelvin. Numbers mark energy minima on the potential
energy surface. Arrows show the transition paths between different
conformations of the molecule.} \label{map_gly3}
\end{figure}

\begin{figure}[h]
\includegraphics[scale=0.7,clip]{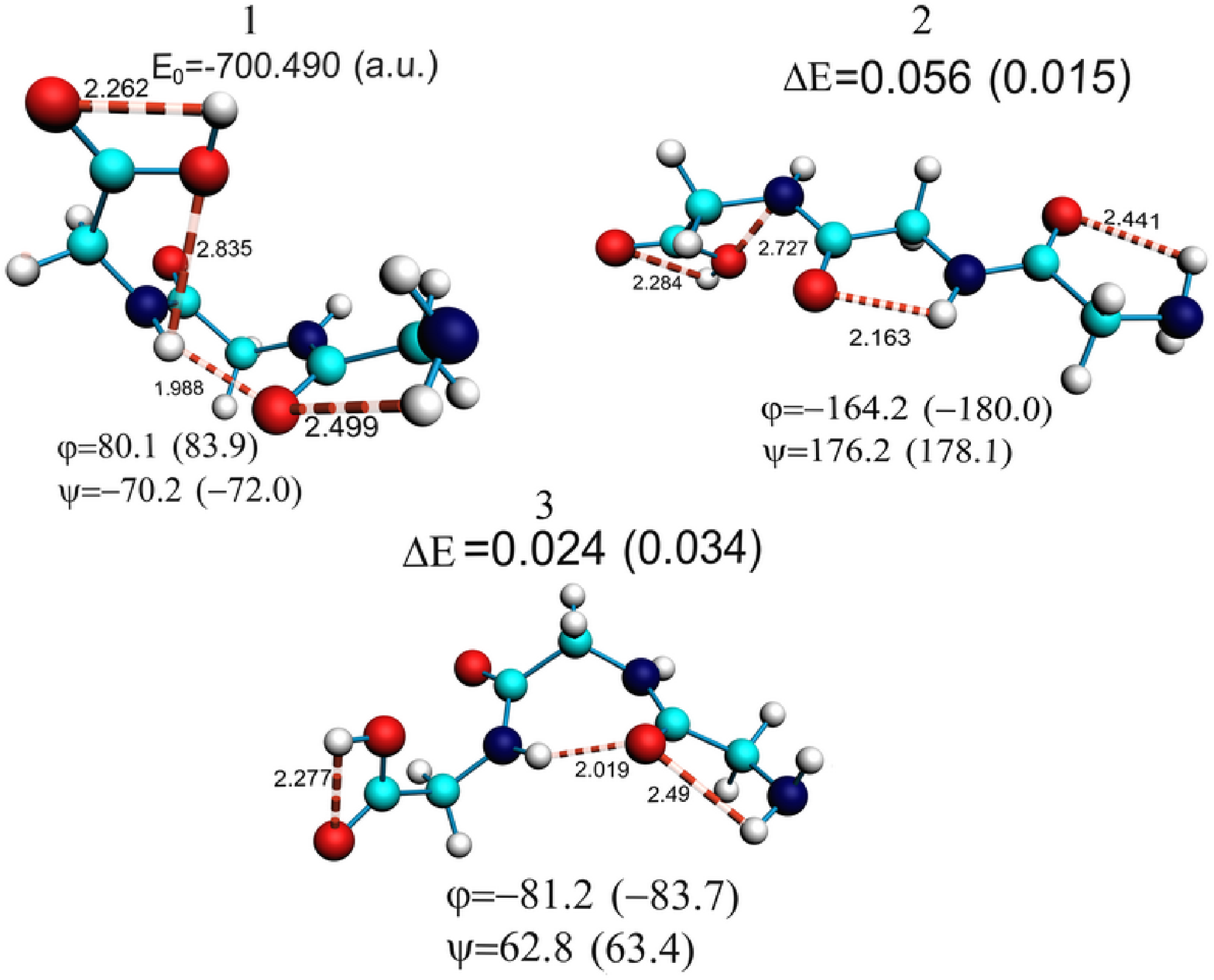}
\caption{Optimized conformations of the glycine tripeptide.
Different geometries correspond to different minima on the potential
energy surface (see contour plot in figure \ref{map_gly3}). Below
each image we present angles $\varphi$ and $\psi$, which have been
obtained with accounting for relaxation of all degrees of freedom in
the system. Values in brackets give the angles calculated without
accounting for relaxation. Above each image, the energy of the
corresponding conformation is given in eV. The energies are counted
from the energy of conformation 1 (the energy of conformation 1 is
given in a.u.). Values in brackets give the energies obtained
without accounting for the relaxation of all degrees of freedom in
the system. Dashed lines show the strongest hydrogen bonds. Their
lengths are given in angstroms.} \label{geom_gly3}
\end{figure}

In figure \ref{map_gly3} we present the potential energy surface for
the glycine tripeptide calculated by the B3LYP/6-31G(2d,p) method.
The energy scale is given in eV, kcal/mol and Kelvin. Energies on
the plot are measured from the lowest energy minimum of the
potential energy surface.

From the figure follows that there are several minima on the
potential energy surface. They are numbered according to the value
of the corresponding energy value. Each minimum corresponds to a
certain conformation of the molecule. These conformations differ
significantly from each other. In the case of glycine tripeptide
there are only three conformations, shown in figure \ref{geom_gly3}.
Dashed lines show the strongest hydrogen bonds in the system, which
arise when the distance between hydrogen and oxygen atoms becomes
less then 2.9 angstroms.

To calculate the potential energy surface the following procedure
was adopted. Once the stable structure of the molecule has been
determined and optimized, all but two (these are the angles
$\varphi$ and $\psi$ in the central amino acid) degrees of freedom
were frozen. Then the energy of the molecule was calculated by
varying $\varphi$ and $\psi$. This procedure was used to calculate
all potential energy surfaces presented below in this section. It
allows one to find efficiently the minima on the potential energy
surface and to determine the main stable conformations of the
molecule. The absolute energy values of different conformations of
the tripeptide found by this method are not too accurate, because
the method does not account for the relaxation of other degrees of
freedom in the system. To calculate the potential energy surface
with accounting for the relaxation one needs 20-30 times more of the
computer time. Therefore, a systematic calculations with accounting
for the relaxation have not been performed in our work. Instead, we
have performed a complete optimization of the molecular
conformations, corresponding to all minima on the calculated
potential energy surface.

In figure \ref{geom_gly3} we compare stable conformations of the
glycine tripeptide calculated with and without accounting for the
relaxation of all atoms in the system. As it is seen from this
figure the angles $\varphi$ and $\psi$ differ by about 10 percent in
the two cases. This difference arises due to the coupling of
$\varphi$ and $\psi$ with other degrees of freedom. Note the change
of the sign of the relative energies of some conformations. This
effect is due to the rearrangement of side atoms (radicals) in the
polypeptide chain which lowers the energies of different
conformations differently.

In our work the potential energy surface has been calculated and
interpolated on the grid with the step of $18^{\circ}$. This step
size is an optimal one, because the interpolation error is about
$9^{\circ}$, i.e. comparable with the angle deviations caused by the
relaxation of all degrees of freedom in the system.

\begingroup
\begin{table*}[h]
\caption{Comparison of dihedral angles $\varphi$ and $\psi$
corresponding to different conformations of glycine tripeptide
(column 3) with angles $\varphi$ and $\psi$ for glycine dipeptide
from \cite{Head-Gordon91,Gould94} (column 1 and 2).}
\label{tab:Gly3}

\begin{ruledtabular}
\begin{tabular}{c|cc|cc|cc}

     \multicolumn{1}{c|}{ conformation } &
     \multicolumn{1}{c}{$\varphi$, \cite{Head-Gordon91}} &
     \multicolumn{1}{c|}{$\psi$\cite{Head-Gordon91}} &
     \multicolumn{1}{c}{$\varphi$\cite{Gould94}} &
     \multicolumn{1}{c|}{$\psi$\cite{Gould94}}&
     \multicolumn{1}{c}{$\varphi$}&
     \multicolumn{1}{c}{$\psi$}\\

\hline
1 & -      & -     &  76.0  & -55.4 & 80.1   & -70.2 \\
2 & -180.0 & 180.0 & -157.2 & 159.8 & -164.2 & 176.2 \\
3 & -85.2  & 67.4  & -85.8  & 79.0  & -81.2  & 62.8  \\

\end{tabular}
\end{ruledtabular}
\end{table*}
\endgroup

In refs. \cite{Head-Gordon91} and \cite{Gould94} several stable
conformations were found for alanine and glycine dipeptides. The
values of angles $\varphi$ and $\psi$ for the stable conformations
of dipeptide and tripeptide are close indicating that the third
amino acid in tripeptide makes relatively small influence on the
values of dihedral angles of two other amino acids. In earlier
papers \cite{Head-Gordon91,Gould94} dipeptides were studied within
the framework of the Hartree-Fock theory. In ref.
\cite{Head-Gordon91}, values of $\varphi$ and $\psi$ were obtained
by the HF/6-31+G* method, and in ref. \cite{Gould94} by HF/6-31G**.
In table \ref{tab:Gly3} we compare the results of our calculation
for tripeptide with the corresponding data obtained for dipeptides.
Some discrepancy between the values presented is due to the
difference between the dipeptide and tripeptide (i.e. the third
glycine in tripeptide affects the values of angles $\varphi$ and
$\psi$). However, another source of discrepancy might arise due to
accounting for the many-electron correlations in the DFT and
neglecting this effect in the Hartree-Fock theory used in refs.
\cite{Head-Gordon91,Gould94}.

Figure \ref{map_gly3} shows that some domains of the potential
energy surface, where the potential energy of the molecule increases
significantly, appear to be unfavorable for the formation of a
stable molecular configuration. The growth of energy takes place
when some atoms in the polypeptide chain approach each other of
small distances. Accounting for the molecule relaxation results in
the decrease of the system energy in such cases, but the resulting
molecular configurations remain unstable. We call such domains on
the potential energy surface as forbidden ones. In figure
\ref{map_gly3} one can identify two forbidden regions in the
vicinity of the points (0, 0) and (0, 180). At (0, 0) a pair of
hydrogen and oxygen atoms approach to the distances much smaller
than the characteristic $H-O$ bond length. This leads to a strong
interatomic repulsion caused by the exchange interaction of
electrons. At (0, 180) the Coulomb repulsion of pair of oxygen atoms
causes the similar effect.

\begin{figure}[t]
\includegraphics[scale=1.0,clip]{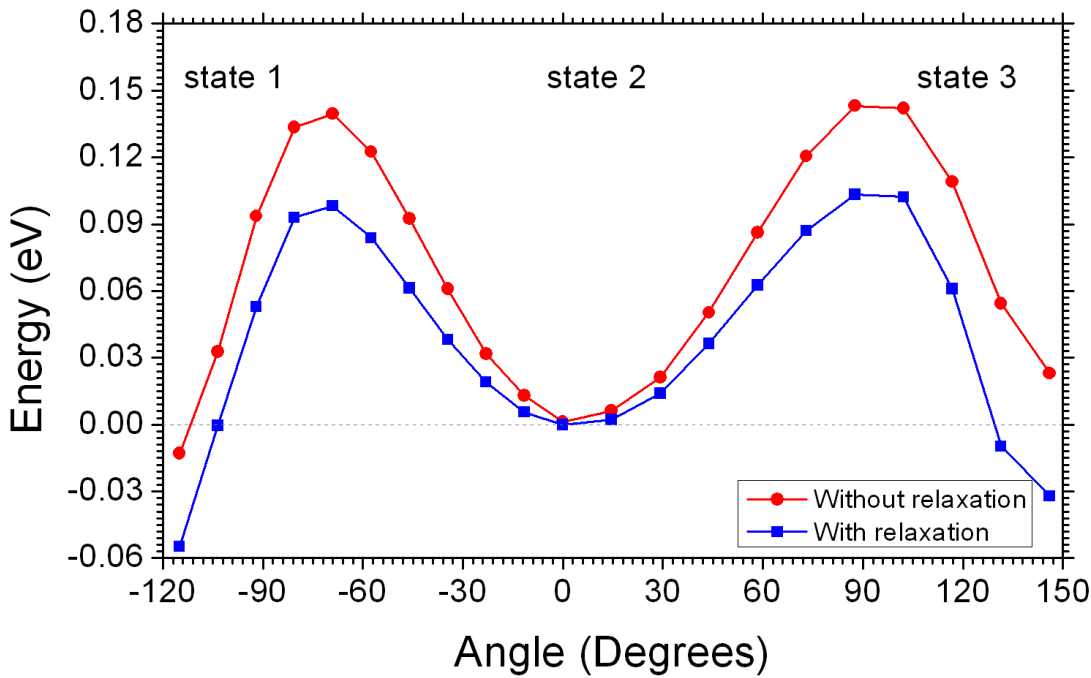}
\caption{Transition barriers between conformations $1
\leftrightarrow 2 \leftrightarrow 3$  of the glycine tripeptide.
Circles and squares correspond to the barriers calculated without
and with relaxation of all degrees of freedom in the system.}
\label{barrier_gly3}
\end{figure}

Figure \ref{map_gly3} shows that there are three minima on the
potential energy surface for glycine tripeptide. The transition
barriers between the conformations $2 \leftrightarrow 1$ and $2
\leftrightarrow 3$ are shown in figure \ref{barrier_gly3}. They have
been calculated with and without relaxation of the atoms in the
system. The corresponding transition paths are marked in figure
\ref{map_gly3} by arrows. This comparison demonstrates that
accounting for the relaxation significantly lowers the barrier
height and influences the relative value of energy of the minima.

\begin{figure}[h]
\includegraphics[scale=1.0,clip]{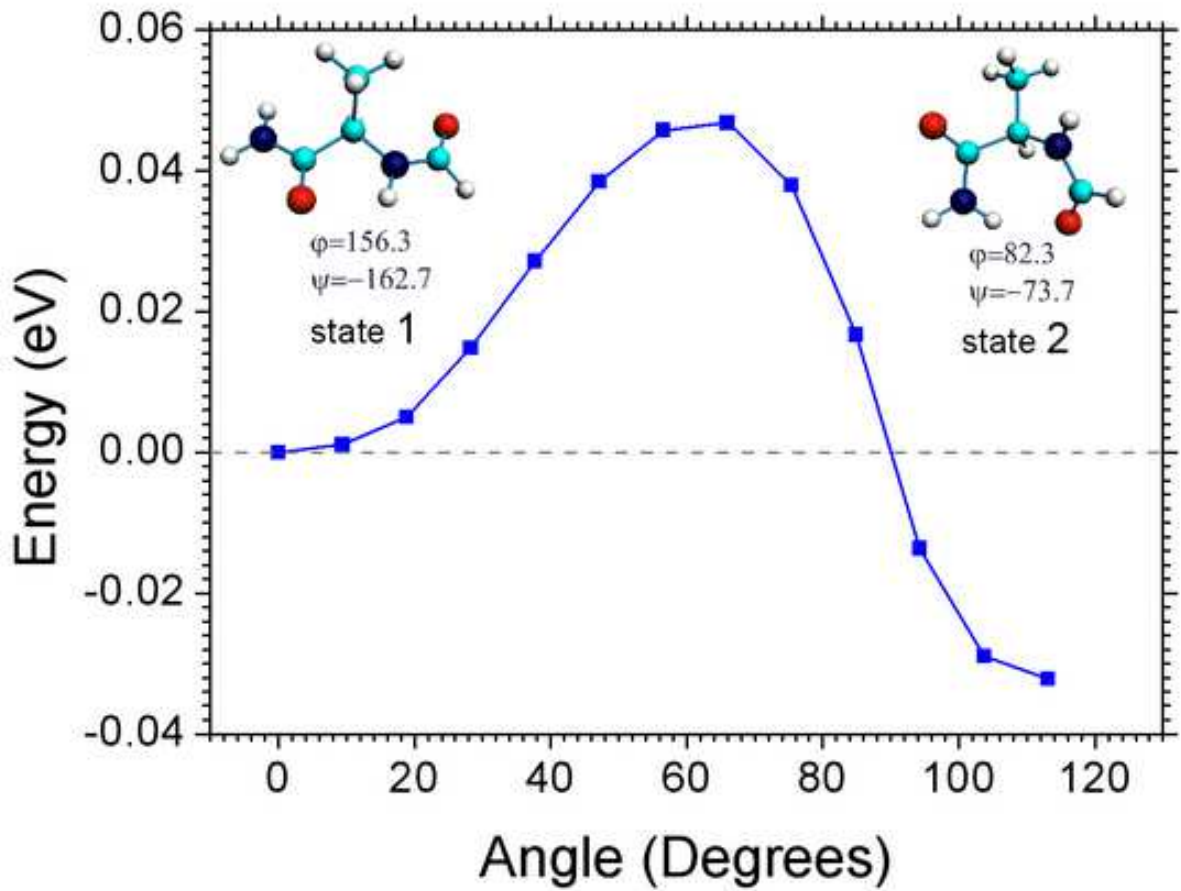}
\caption{Transition barriers between conformations $1
\leftrightarrow 2$ of alanine dipeptide analog calculated by the
B3LYP/6-31+G(2d,p) method accounting for the relaxation of all
degrees of freedom in the system. Structure of the conformations 1
and 2 is shown near each minimum.} \label{barrier_gly2}
\end{figure}

Let us now estimate the time needed for a system for the transition
from one conformation to another. To do this we use the Arhenius
equation, which reads as:

\begin{equation}
\frac{1}{\tau}=\Omega e^{-\frac{\Delta E}{kT}}
\label{Arhenius_eq}
\end{equation}

\noindent where $\tau$ is the transition time, $\Omega$ is the
factor, determining how frequently the system approaches the
barrier, $\Delta E$ is the barrier height, $T$  is the temperature
of the system, $k$  is the Bolzmann factor.

Figure \ref{barrier_gly2} shows the barrier for the transition
between the two main conformations of the alanine dipeptide analog
((S)-$\alpha$-(formylamino)propanamide). It is seen that $\Delta
E_{1\rightarrow2}=0.047$ eV for the transition $1\rightarrow2$,
while $\Delta E_{2\rightarrow1}=0.079$ eV for the transition
$2\rightarrow1$. The frequency $\Omega$ for this molecule is equal
to 42.87 cm$^{-1}$. Thus, at $T=300$ K, we obtain $\tau_{2\times
Ala}^{1\rightarrow2} \sim 5$ ps and $\tau_{2\times
Ala}^{2\rightarrow1} \sim 17$ ps. This result is in excellent
agreement with the molecular dynamics simulations results obtained
in \cite{Salahub01} predicting $\tau\sim7$ ps for the transition
${1\rightarrow2}$ and $\tau\sim19$ ps for the transition
${2\rightarrow1}$. This comparison demonstrates that our method is
reliable enough and it can be used for the estimation of transition
times between various conformations of the polypeptides.

Using the B3LYP/6-31G(2d,p) method we have calculated the
frequencies of normal vibration modes for the glycine tripeptide.
The characteristic frequency corresponding to twisting of the
polypeptide chain is equal to 33.49 cm$^{-1}$. From figure
\ref{barrier_gly3} follows that $\Delta E_{2\rightarrow3}=0.103$ eV
for the transition $2\rightarrow3$, and $\Delta
E_{3\rightarrow2}=0.132$ eV for the transition $3\rightarrow2$.
Thus, we obtain $\tau_{3\times Gly}^{2\rightarrow3} \sim 54$ ps and
$\tau_{3\times Gly}^{3\rightarrow2} \sim 164$ ps. Let us note that
these transition times can be measured experimentally by means of
NMR \cite{Rubin04,Bax03}.

\subsection{Potential energy surface for glycine hexapeptide with
the sheet and the helix secondary structure}

\begin{figure}[h]
\includegraphics[scale=0.835,clip]{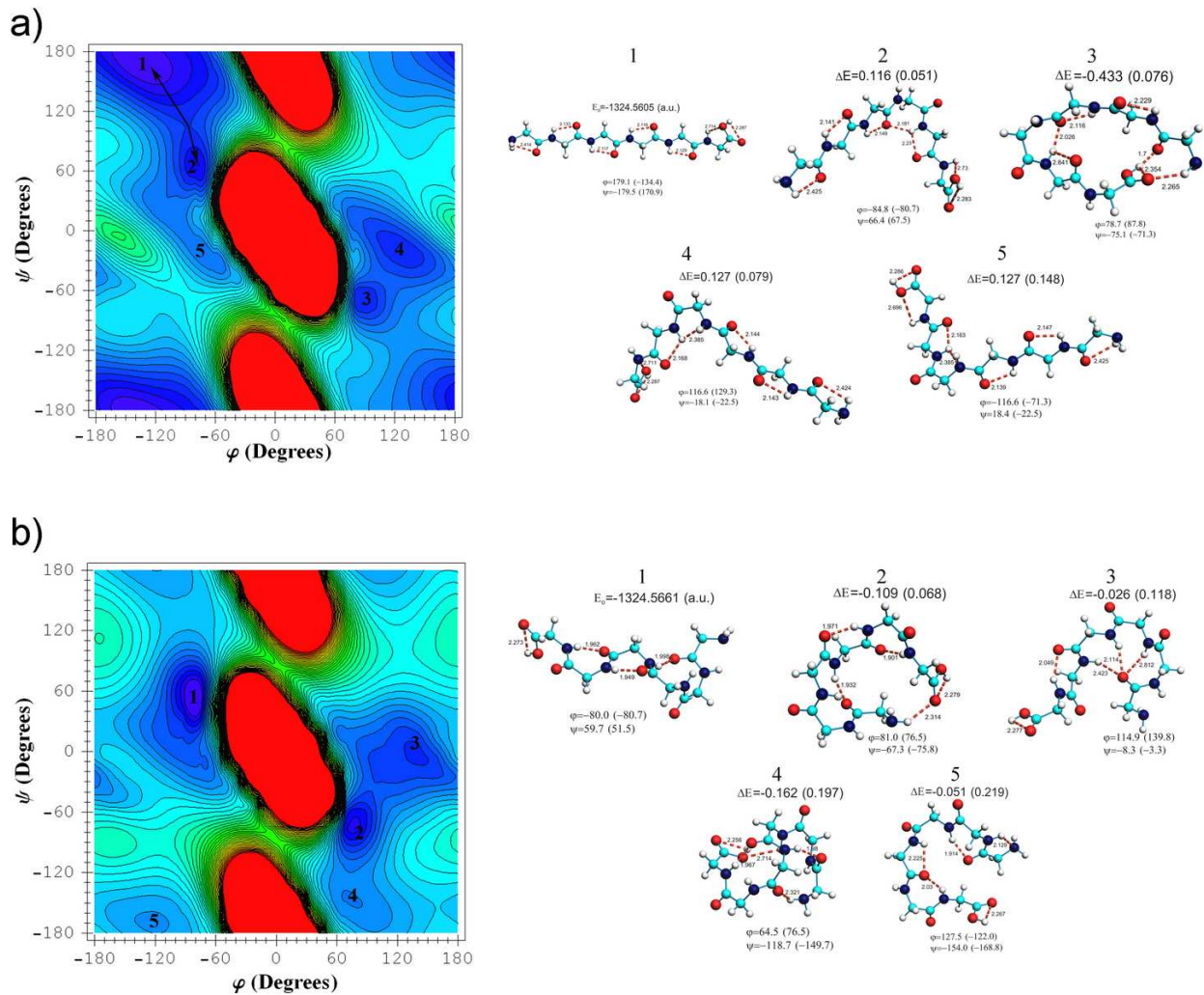}
\caption{Potential energy surface for the glycine hexapeptide with
the sheet secondary structure (part a) and with the helix secondary
structure (part b) calculated by the B3LYP/6-31G(2d,p) method.
Energy scale is given in figure \ref{map_gly3}. Numbers mark energy
minima on the potential energy surface. Images of optimized
conformations of the glycine hexapeptide are shown near the
corresponding energy landscape. Values of angles $\varphi$ and
$\psi$, as well as the relative energies of the conformations are
given analogously to that in figure \ref{geom_gly3}.} \label{gly6}
\end{figure}

In figure \ref{gly6} we present contour plots of the potential
energy surfaces for the glycine hexapeptide with the sheet (part a)
and the helix (part b) secondary structure, respectively, versus
dihedral angles $\varphi$ and $\psi$. In both cases the forbidden
regions arise because of the repulsion of oxygen and hydrogen atoms
analogously to the glycine tripeptide case.

Minima 1-5 on the potential energy surface \ref{gly6}a correspond to
different conformations of the glycine hexapeptide with the sheet
secondary structure. Note that minima 1-3 are also present on the
potential energy surface of the glycine tripeptide. Geometries of
the conformations 1-5 are shown on the right-hand side of figure
\ref{gly6}a.

\begin{figure}[h]
\includegraphics[scale=0.7,clip]{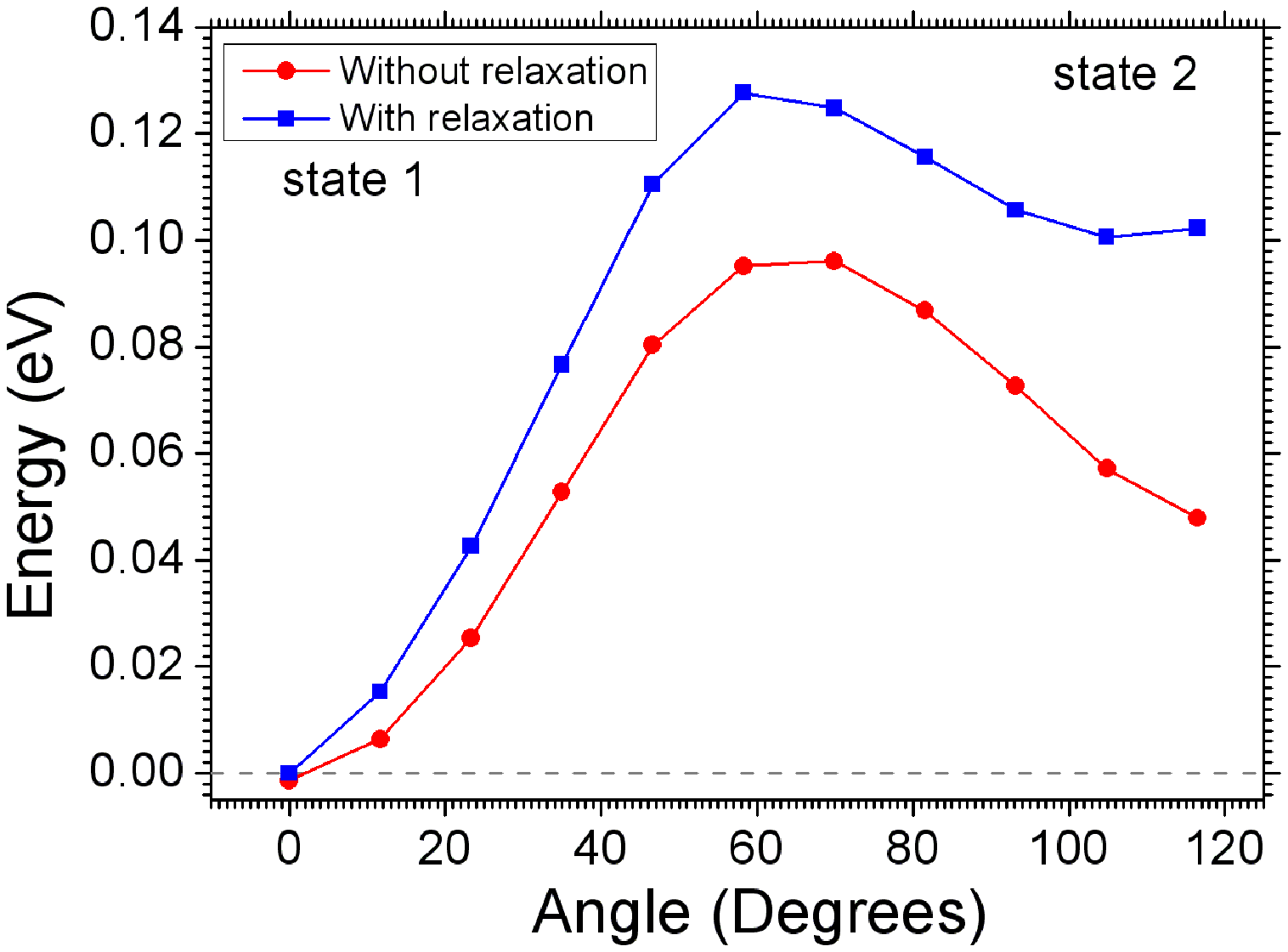}
\caption{Transition barriers between conformations $1
\leftrightarrow 2$ of the glycine hexapeptide with the sheet
secondary structure. Circles and squares correspond to the barriers
calculated without and with relaxation of all degrees of freedom in
the system.} \label{barrier_gly6_sheet}
\end{figure}

For the glycine hexapeptide with the sheet secondary structure
additional minima 4-5 arise. The appearance of these minima is the
result of the interaction of the outermost amino acids, which are
absent in the case of tripeptide.

Energy barrier as a function of a scan variable (see figure
\ref{gly6}a) for the transition between conformations 1 and 2 is
shown in figure \ref{barrier_gly6_sheet}. The energy dependence has
been calculated with and without relaxation of all the atoms in the
system. In the case of glycine hexapeptide with the sheet secondary
structure the barrier height (0.128 eV) for the transition
$1\rightarrow2$ appears to be close to the corresponding barrier
height of the glycine tripeptide (0.103 eV), while the barrier
height for the transition $2\rightarrow1$ is significantly lower
(0.028 eV). The normal vibration mode frequency, corresponding to
the twisting of the polypeptide chain is equal to 15.45 cm$^{-1}$
and was calculated with the B3LYP/6-31G(2d,p) method. Using equation
(\ref{Arhenius_eq}) one derives the transition times at room
temperature: $\tau_{6\times Gly}^{1\rightarrow2} \sim 305$ ps,
$\tau_{6\times Gly}^{2\rightarrow1} \sim 6\ $ps.

Let us now consider glycine hexapeptide with the helix secondary
structure. The potential energy surface for this polypeptide is
shown in figure \ref{gly6}b. The positions of minima on this surface
are shifted significantly compared to the cases discussed above.
This change takes place because of the influence of the secondary
structure of the polypeptide on the potential energy surface. The
geometries of the most stable conformations are shown on the right
hand-side of figure \ref{gly6}b.

It is worth noting that for some conformations of glycine
hexapeptide the angles $ \varphi $ and $ \psi $ change significantly
when the relaxation of all degrees of freedom in the system is
accounted for (see for example conformations 1, 4, 5 in fig.
\ref{gly6}a and conformations 3, 4 in fig. \ref {gly6}b). This means
that the potential energy surface of the glycine hexapeptide in the
vicinity of mentioned minima is very sensitive to the relaxation of
all degrees of freedom. However, calculation of the potential energy
surface with accounting for the relaxation of all degrees of freedom
is unfeasible task. Indeed, one needs about 1000 hours of computer
time (Pentium Xeon 2.4 GHz) for the calculation of the potential
energy surface for the glycine hexapeptide. To perform an analogues
calculation with accounting for the relaxation about 3 years of
computer time would be needed. Nevertheless, the potential energy
surface calculated without accounting for the relaxation carries a
lot of useful information. Thus, one can predetermine stable
conformations of polypeptide, which then can be used as starting
configurations for further energy minimization.

\subsection{Comparison of calculation results with experimental data}

Nowadays, the structure of many proteins has been determined
experimentally \cite{Protbase}. Knowing the protein structure one
can find the angles $\varphi$ and $\psi$ for each amino acid in the
protein.

\begin{figure}[p]
\includegraphics[scale=0.85,clip]{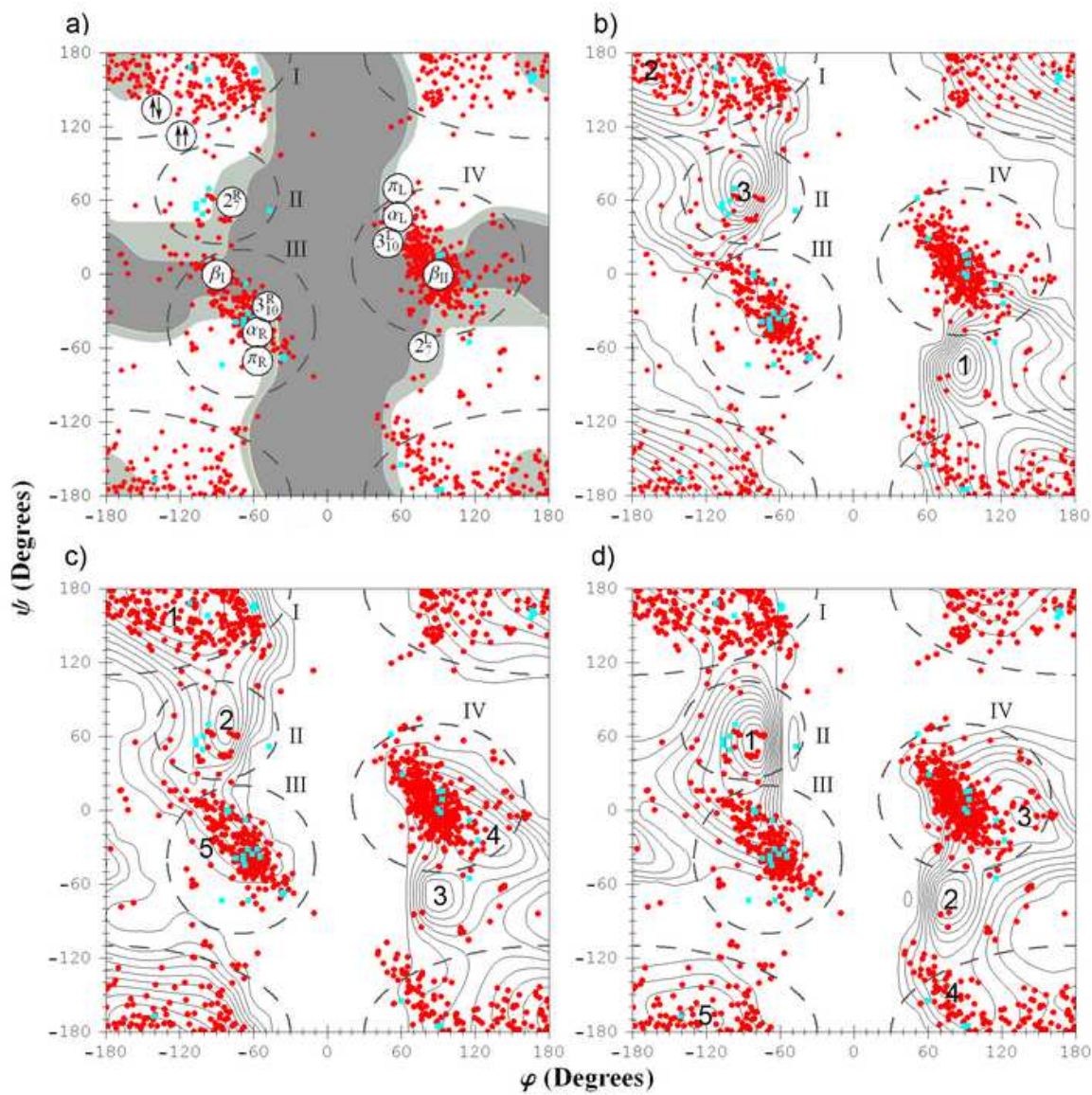}
\caption{Comparison of angles $\varphi$ and $\psi$ of glycine
residues in protein structures selected from the Brookhaven Protein
Data Bank \cite{Protbase,RamachandranPlot} with the steric diagram
for poly-glycine \cite{Biochemistry_book} (part a)). Comparison of
angles $\varphi$ and $\psi$ of glycine residues in protein
structures selected from the Brookhaven Protein Data Bank
\cite{Protbase,RamachandranPlot} with the minima on the calculated
potential energy surfaces for: glycine tripeptide (b); glycine
hexapeptide in sheet conformation (c); glycine hexapeptide in helix
conformation (d). Transparent rhomboids correspond to glycines
surrounded with glycines, while filled circles correspond to
glycines surrounded by other amino acids. Dashed ellipses mark the
regions of higher concentration of the observed angles.}
\label{Gly_experiment}
\end{figure}

In figure \ref{Gly_experiment}a, we show a map of the allowed and
forbidden conformations for glycine residues in poly-glycine chain
taken from \cite{Biochemistry_book} (steric Ramachandran diagram).
This map was obtained from pure geometrical considerations, in which
the structure of the polypeptide was assumed to be fixed and defined
by the interatomic van der Waals interaction radii. Depending on the
distances between the atoms one could distinguish three regions:
completely allowed, conventionally allowed and forbidden. The
conformation is called completely allowed if all the distances
between atoms of different amino acids are larger than some critical
value $r_{ij}\ge r_{max}$. Conventionally allowed regions on the
potential energy surface correspond to the conformations of the
polypeptide, in which the distances between some atoms of different
amino acids lie within the interval $r_{min}\le r_{ij} < r_{max}$.
All other conformations are referred to as forbidden. The values of
$r_{min}$ and $r_{max}$ are defined by the types of interacting
atoms and can be found in the textbooks (see, e.g.,
\cite{Biochemistry_book}). In figure \ref{Gly_experiment}a we mark
the completely allowed regions with white, the conventionally
allowed regions with light-gray and the forbidden regions with dark
gray color. In this figure we mark the points, which correspond to
the geometries of glycine, whose periodical iteration leads to the
formation of chains with specific secondary structure. In table
\ref{tab:RamAngl} we compile the values of angles $\varphi$ and
$\psi$, which correspond to the most prominent poly-glycine
secondary structures. For the illustrative purposes we mark these
points by white circles with the corresponding type of the secondary
structure typed in. Thus, $2_7^{R}$, $2_7^{L}$ are the right-handed
and the left-handed $2_7$ helix; $3_{10}^{R}$, $3_{10}^{L}$ are the
right-handed and the left-handed $3_{10}$ helix; $\alpha_{R}$,
$\alpha_{L}$ are the right-handed and the left-handed $\alpha-$helix
($4_{13}$); $\pi_{R}$, $\pi_{L}$ are the right-handed and the
left-handed $\pi-$helix ($5_{16}$); $\uparrow\uparrow$,
$\uparrow\downarrow$ are the parallel and antiparallel $\beta$
sheets. $\beta_{I}$, $\beta_{II}$ correspond to the $\beta-$turns of
types I and II respectively.

\begingroup
\begin{table*}[h]
\caption{Angles $\varphi$ and $\psi$ corresponding to the most
prominent poly-glycine secondary structures.} \label{tab:RamAngl}

\begin{ruledtabular}
\begin{tabular}{ccc}

     \multicolumn{1}{c}{Structure type} &
     \multicolumn{1}{c}{$\varphi$ (Deg.)} &
     \multicolumn{1}{c}{$\psi$ (Deg.)}\\

\hline
right-handed (left-handed) $2_7$ helix    & -78 (78) & 59 (-59)  \\
right-handed (left-handed) $3_{10}$ helix & -49 (49) & -26 (26)  \\
right-handed (left-handed) $\alpha-$helix ($4_{13}$) & -57 (57) & -47 (47)  \\
right-handed (left-handed) $\pi-$helix ($5_{16}$)   & -57 (57) & -70 (70)  \\
parallel $\beta$ sheet ($\uparrow\uparrow$) & -119 & 113 \\
antiparallel $\beta$ sheet ($\uparrow\downarrow$) & -139 & 135 \\
$\beta-$turn of type I & -90 & 0 \\
$\beta-$turn of type II & 90 & 0 \\

\end{tabular}
\end{ruledtabular}
\end{table*}
\endgroup

Note that not all of the structures listed above are present equally
in proteins. In figure \ref{Gly_experiment}a we show the
distribution of the angles $\varphi$ and $\psi$ of glycine residues
in protein structures selected from the Brookhaven Protein Data Bank
\cite{Protbase,RamachandranPlot}. It is possible to distinguish four
main regions, in which most of the experimental points are located.
In figure \ref{Gly_experiment} these regions are schematically shown
with dashed ellipses. Note that these ellipses are used for
illustrative purposes only, and serve for a better understanding of
the experimental data. The regions in which most of the observed
angles $\varphi$ and $\psi$ are located correspond to different
secondary structure of the poly-glycine. Thus, region I corresponds
to the parallel and antiparallel $\beta-$sheets. Region II
corresponds to the right-handed $2_7^R$ helix. Region III
corresponds to the right-handed $\alpha_R-$helix, right-handed
$3_{10}^R$ helix, right-handed $\pi_R-$helix and $\beta-$turn of
type I. Region IV corresponds to the left-handed $\alpha_L-$helix,
left-handed $3_{10}^L$ helix and $\beta-$turn of type II. In some
cases there are several types of secondary structure within one
domain. In the present work we have not studied the secondary
structure of proteins systematically enough to establish the
univocal correspondence of the observed experimental points to
different types of the secondary structure. Figure
\ref{Gly_experiment}a demonstrates that $\pi_{L}$ and $2_7^{L}$
helixes are rarely observed in experiment.

Note that some experimental points lie in the forbidden region of the
steric Ramachandran diagram (see region IV in fig. \ref{Gly_experiment}a).
The quantum calculation shows that in fact this region is allowed
and has several minima on the potential energy surface (see fig \ref{Gly_experiment}c
and \ref{Gly_experiment}d). This comparison shows that for an accurate description
of polypeptides it is important to take quantum properties of these
systems into account.

Let us now compare the distribution of angles $\varphi$ and $\psi$
experimentally observed for proteins with the potential energy
landscape calculated for glycine polypeptides, and establish
correspondence of the secondary structure of the calculated
conformations with the predictions of the simple Ramachandran model.

Region I corresponds to the minima 2, 1 and 5 on the potential
energy surfaces of the glycine tripeptide (fig.
\ref{Gly_experiment}b), the glycine hexpeptide with the secondary
structure of sheet (fig. \ref{Gly_experiment}c), and the glycine
hexapeptide with the secondary structure of helix (fig.
\ref{Gly_experiment}d) respectively. Conformations 2 and 1 in
figures \ref{Gly_experiment}b and \ref{Gly_experiment}c correspond
exactly to the glycine chains in the $\beta$-sheet conformation (see
fig. \ref{geom_gly3} and \ref{gly6}a). Conformation 5 in figure
\ref{Gly_experiment}d is a mixed state. Here the central amino acid
has the conformation of sheet, while the outermost amino acids have
the conformation of helix.

Region II corresponds to the minima 3, 2 and 1 on the potential
energy surfaces \ref{Gly_experiment}b, \ref{Gly_experiment}c, and
\ref{Gly_experiment}d, respectively. On the steric diagram for
poly-glycine (see fig. \ref{Gly_experiment}a) this region
corresponds to the right-handed $2_7^R$ helix. The structure of
conformations 3 and 2 on the surfaces \ref{Gly_experiment}b and
\ref{Gly_experiment}c differs from the structure of this particular
helix type. Only the central glycines, for which the angles
$\varphi$ and $\psi$ in figures \ref{Gly_experiment}b and
\ref{Gly_experiment}c are defined, have the structure of $2_7^R$
helix. Thus, one can refer to conformations 3 and 2 as to the mixed
states, where the central part of the polypeptide chain has the
conformation of helix and the outermost parts have the conformation
of sheet. Conformation 1 on the surface \ref{Gly_experiment}d is
also a mixed state. Here one can distinguish one turn of $3_{10}^R$
helix and two turns of $2_7^R$ helix (see fig. \ref{gly6}b).

Region III corresponds to the structure of right-handed
$\alpha_R-$helix, right-handed $3_{10}^R$ helix, right-handed
$\pi_R-$helix and  $\beta-$turn I. It corresponds to minimum 5 on
the potential energy surface of the glycine hexapeptide with the
secondary structure of sheet \ref{Gly_experiment}c. Conformation 5
can be characterized as partially formed $\beta-$turn, because the
glycine, for which the dihedral angles $\varphi$ and $\psi$ in
figure (fig. \ref{Gly_experiment}c) are defined, has the geometry of
$\beta-$turn and its neighbor forms a $\beta-$sheet (see fig.
\ref{gly6}a). There are no minima in region III on the potential
energy surfaces presented in figures \ref{Gly_experiment}b and
\ref{Gly_experiment}d. This happens because in this case most
probable is the structure of right-handed $\alpha_R-$helix. To form
one turn of the helix of this type it is necessary to link at least
four amino acids, so the glycine tripeptide is too short for that.
As well, six amino acids chain is too short to form a stable
fragment of an $\alpha_R-$helix, because it does not have enough
hydrogen bonds to stabilize the structure. For the hexapeptide more
probable are the elements of $3_{10}^R$ and $2_{7}^R$ helixes,
because in these cases 2 and 3 helical turns respectively can be
formed.

Region IV is represented by the structure of left-handed
$\alpha_L-$helix, left-handed $3_{10}^L$ helix and $\beta-$turn of
type II. To form these structures it is necessary to have at least
four amino acids, therefore there is no minima in this region on the
potential energy surface of glycine tripeptide (fig.
\ref{Gly_experiment}b). Region IV corresponds to the conformations 4
and 3 on the surfaces \ref{Gly_experiment}c and
\ref{Gly_experiment}d respectively. Conformation 4 on the surface
\ref{Gly_experiment}c corresponds to partially formed $\beta-$turn,
because the glycine, for which the dihedral angles $\varphi$ and
$\psi$ in figure \ref{Gly_experiment}c are plotted has the
configuration of $\beta-$turn, but the neighboring aminoacid in the
chain forms a $\beta-$sheet (see fig. \ref{gly6}a). Conformation 3
on the surface \ref{Gly_experiment}d can be characterized as
deformed turn of left-handed $\alpha_L-$helix, which turns out to be
energetically the most favorable in this region of the potential
energy surface (see fig. \ref{gly6}b).

Finally, we mention a few peculiarities of the calculated potential
energy landscapes. At each of the potential energy surfaces
discussed in this work one can see a minimum at $\varphi\sim80^{o}$
and $\psi\sim-70^{o}$. On the steric diagram for poly-glycine this
region corresponds to the left-handed $2_{7}^L$ helix. Conformations
1, 3 and 2 on the potential energy surfaces \ref{Gly_experiment}b,
\ref{Gly_experiment}c, and \ref{Gly_experiment}d respectively
partially represent this structure (see fig. \ref{geom_gly3},
\ref{gly6}a and \ref{gly6}b). The structure of conformation 4 on the
potential energy surface \ref{Gly_experiment}d is similar to
left-handed $\alpha_L-$helix, but differs from it due to the short
length of the polypeptide chain, resulting in significant variation
of angles $\varphi$ and $\psi$ in all the residues along the chain.

\section{Conclusion}
\label{conclusions}

In the present paper the multidimensional potential energy surfaces
for amino acid chains consisting of three and six glycines has been
investigated and the conformational properties of these systems with
respect to the twisting of the polypeptide chain have been
described. The calculations were carried out within {\it ab initio}
theoretical framework based on the density functional theory
accounting for all the electrons in the system. We have determined
stable conformations and calculated the energy barriers for
transitions between them. Using a thermodynamic approach, we have
estimated times of the characteristic transitions between the
conformations. It was demonstrated that the transition times lie
within the picosecond region. Our estimates are compared with the
available molecular-dynamics simulations results, and the
correspondence between the results of the two different methods is
reported. A strong barrier asymmetry between neighboring stable
conformations on the potential energy surface was found.

We have compared for the first time values of angles $\varphi$ and
$\psi$ for glycine residues experimentally observed in real proteins
with the coordinates of minima on the potential energy surfaces.
This comparison has showed that all profound minima on the potential
energy surfaces correspond to the regions in which experimentally
measured values of $\varphi$ and $\psi$ are located. We have also
analysed how the secondary structure of polypeptide chains
influences the formation of the potential energy landscapes. For the
chains of six amino acids with the secondary structures of sheet and
helix the influence of the secondary structure on the stable
conformations of the molecule was demonstrated.

The results of this work can be utilized for modeling more complex
molecular systems. For example, the suggested model for the
estimation of the characteristic transition times can be used for
longer polypeptide chains, also consisting of different amino acids
and for estimates of time of proteins folding. It is also possible
to use the results of the present work for testing the applicability
and accuracy of different model approaches for the polypeptide
description requiring much less computer time than {\it ab initio}
calculations.

\section{Acknowledgements}
This work is partially supported by the European Commission within
the Network of Excellence project EXCELL, by INTAS under the grant
03-51-6170 and by the Russian Foundation for Basic Research under
the grant 06-02-17227-a. We are grateful to Dr. A. Korol and Dr. O.
Obolensky for their help in preparation of this manuscript. The
possibility to perform complex computer simulations at the Frankfurt
Center for Scientific Computing is also gratefully acknowledged.

\end{document}